\documentclass[a4paper, 12pt]{article}
\author{Clovis Jacinto de Matos\footnote{ESA-HQ, European Space Agency, 8-10 rue Mario Nikis, 75015 Paris, France, e-mail: Clovis.de.Matos@esa.int}
}
\title{Electromagnetic Dark Energy and Gravitoelectrodynamics of Superconductors}

\begin{document}

\maketitle \begin{abstract}It is shown that Beck and Mackey
electromagnetic model of dark energy in superconductors can
account for the non-classical inertial properties of
superconductors, which have been conjectured by the author to
explain the Cooper pair's mass excess reported by Cabrera and
Tate. A new Einstein-Planck regime for gravitation in condensed
matter is proposed as a natural scale to host the
gravitoelectrodynamic properties of superconductors.
\end{abstract}

\section{Introduction}
We start from the unsolved problem of the Cooper pairs excess of
mass. A conjecture involving a gravitomagnetic London moment in
superconductors to account for this anomaly is reviewed.
Afterwards the relation of this new phenomena in superconductors
with the spontaneous breaking of the principle of general
covariance and its relation with the problem of the graviton mass
and of dark energy is briefly discussed. The link with Beck's
electromagnetic model for dark energy in superconductors is then
established. Assuming that different superconductive materials
host different vacuum energy densities allows to account very
accurately for the conjectured gravitomagnetic fields in
superconductors. It appears that this new phenomena takes place at
the Einstein-Planck scale defined from the five fundamental
constants of nature: $c, \hbar, G, k, \Lambda$. Unexpectedly this
seems to be the natural scale for an intermediate regime of
quantum-gravity.

\section{Cooper pairs mass excess}
In 1989 Cabrera and Tate, through the measurement of the London
moment, reported an anomalous Cooper pair mass excess in thin
rotating Niobium superconductive rings.
\begin{equation}
\frac{m^*-m}{m}=\frac{\Delta m}{m}=9.2\times10^{-5}\label{1}
\end{equation}
where $m^*=1.82203\times10^{-30} Kg$ is the Cooper Pair mass
experimentally measured, and $m=1.82186\times10^{-30} Kg$ is the
theoretical mass of Cooper pairs including relativistic
corrections \cite{Tate89}\cite{Tate90}.

In an attempt to explain this anomalous excess of mass the author
conjectured that the Cooper pairs mass do not increase but that
instead an additional gravitomagnetic London-type moment must be
taken into account in the quantization of the Cooper pairs
canonical momentum \cite{Tajmar03}\cite{Tajmar05}.
\begin{equation}
B_g=\frac{\Delta m}{m}2 \omega=1.84\times10^{-4}\omega\label{2}
\end{equation}
Where $\omega$ is the superconductor's angular velocity and $B_g$
is the gravitomagnetic field both expressed in $Rad/s$. The
Gravitomagnetic London moment received some preliminary
experimental confirmation in dedicated experiments conducted at
the Austrian Research Centres (ARC) in Seibersdorf
\cite{Tajmar1}\cite{Tajmar2}. In the rest of the text we will
refer to the coupling between $B_g$ and $\omega$ in rotating
superconductors as
\begin{equation}
\chi=\frac{B_g}{\omega}=2\frac{\Delta
m}{m}=1.84\times10^{-4}\label{3}
\end{equation}

First steps to understand this anomalous coupling, in the
framework of a spontaneous breaking of the Principle of General
Covariance (PGC) in superconductors, lead to consider a massive
spin one graviton to convey the gravitoelectromagnetic
interaction, and a set of Einstein-Maxwell-Proca equations to
describe the gravitoelectrodynamics of superconductors \cite{de
Matos}\cite{de Matos1}\cite{de Matos2}. Solving these equations
for gravitomagnetic fields we find the gravitomagnetic London
moment expressed as a function of a density of energy $\rho^*$
contained in the superconductive ring, and of the square of the
graviton Compton wavelength $\lambda_g$.
\begin{equation}
\chi=\frac{8\pi G}{c^4}\rho^*\lambda_g^2\label{4}
\end{equation}
Novello and others \cite{Novello1} \cite{Novello2} \cite{Liao}
proposed a link between the Cosmological Constant (CC), $\Lambda$,
and a massive graviton,
\begin{equation}
\frac{1}{\lambda_g^2}=\Big(\frac{m_g
c}{\hbar}\Big)^2=\frac{2}{3}\Lambda\label{5}
\end{equation}
coming as a natural consequence of the equations of motion of a
massive graviton propagating in a de-Sitter background. On the
other side a non-vanishing cosmological constant can be
interpreted in terms of a non-vanishing vacuum energy called dark
energy.
\begin{equation}
\rho_{vac}=\frac{c^4}{8\pi G}\Lambda\label{6}
\end{equation}
The small experimental value of the CC
$\Lambda=1.29\times10^{-52}[1/m^2]$ \cite{Spergel} and its origin
remain a deep mystery. This is often call the CC problem, since at
Planck scale the vacuum energy density should be of the order of
$10^{120}$, in complete contradiction with measured value. Putting
Equ.(\ref{5}) and Equ.(\ref{6}) into Equ.(\ref{4}) and rearranging
we get:
\begin{equation}
\chi=\frac{3}{2}\frac{\rho^*}{\rho_{vac}}\label{7}
\end{equation}
From Equ.(\ref{7}) we formulate the hypothesis that $\rho^*$
corresponds to the density of dark energy contained in a given
superconductor. It should be stressed that this hypothesis
deviates significantly from the initial attempt in \cite{de
Matos1} to understand the gravitomagnetic London moment, in which
$\rho^*$ was considered as being the Cooper pairs mass density
\cite{Modanese}.

\section{Electromagnetic Dark Energy in Superconductors}
To solve the CC problem Beck and Mackey proposed a model of dark
energy based on electromagnetic vacuum fluctuations creating a
small amount of vacuum energy density exactly equal to the
cosmological vacuum energy density. They assume that in a
superconductor virtual photons, with energy
$\epsilon=\frac{1}{2}h\nu$, can exist in two different phases: A
\emph{gravitationally active} phase where they contribute to the
dark energy density, and a \emph{gravitationally inactive} phase
where they do not contribute to the dark energy density
\cite{Beck}\cite{Beck2}\cite{Beck3}\cite{Beck4}. The transition
between the two phases is defined by a cutoff frequency, $\nu_c$.
They constructed a Ginzburg-Landau type theory for the number
density of gravitationally active photons in superconductors. In
this way they obtain a finite dark energy density in
superconducting materials dependent on a cutoff frequency $\nu_c$:
\begin{equation}
\rho^*=\frac{1}{2}\frac{\pi h}{c^3}\nu_c^4\label{8}
\end{equation}
Equaling Equ(\ref{8}), where $\rho^*$ is the dark energy density
in superconductors, to Equ.(\ref{6}), where $\rho_{vac}$ is the
density of dark energy in the Universe, the cosmological cutoff
frequency in superconductors is estimated to be of the order of
$\nu_c\simeq 2.01 THz$. An experimental effort is currently
undergoing at UCL and Cambridge to measure this cutoff frequency
through the measurement of the spectral density of the noise
current in resistively shunted Josephson junctions \cite{Koch}.
The formal attribution of a temperature $T$ to the virtual photons
underlying dark energy is done by comparing their energy to the
one of ordinary photons in a bath at temperature T.
\begin{equation}
\frac{1}{2}h\nu=\frac{h\nu}{e^{\frac{h\nu}{kT}}-1}\label{9}
\end{equation}
This condition is equivalent to
\begin{equation}
h\nu=\ln3 kT\label{10}
\end{equation}
Using Equ.(\ref{10}), the critical temperature $T_c$ in the
Beck-Ginzburg-Landau model corresponds to a critical frequency
\begin{equation}
\nu_c=\ln3 \frac{k T_c}{h}\label{11}
\end{equation}
Putting the cosmological cutoff frequency estimated above,
$\nu_c=2.01 THz$ into Equ.(\ref{11}) we find the associated
critical temperature, $T_c=87.49 K$, which is characteristic of
$High-T_c$ superconductors.

In the present work we consider that the critical temperature,
$T_c$, characterizing a given superconductive material, will
define its respective cutoff frequency, $\nu_c$, through
Equ.(\ref{11}), which in turn will determine the density of dark
energy in the superconductor through Equ.(\ref{8}). $T_c$ being
different for different superconducting materials, $\nu_c(T_c)$
and $\rho^*(T_c)$, will accordingly adopt different values in
different superconducting materials.

\section{The Planck-Einstein Regime of Gravitation}
Substituting Equ.(\ref{6}), Equ.(\ref{8}) and Equ.(\ref{11}) into
Equ.(\ref{7}) we end with:
\begin{equation}
\chi=\frac{3\ln^4 3}{4 \pi}\frac{k^4G}{c^7\hbar^3 \Lambda}
T_c^4\label{12}
\end{equation}
Defining the Planck-Einstein temperature, $T_{PE}$
\begin{equation}
T_{PE}=\frac{1}{k}\Bigg(\frac{c^7\hbar^3
\Lambda}{G}\Bigg)^{1/4}=60.71 K \label{13}
\end{equation}
Equ.(\ref{12}) can be written in the following form:
\begin{equation}
\chi=\frac{3\ln^4 3}{4
\pi}\Bigg(\frac{T_c}{T_{PE}}\Bigg)^4\label{14}
\end{equation}
As indicated above, in the present work, we release the
constraint, with respect to Beck and Mackey initial model, that
all superconductive materials host the same cosmological vacuum
energy density. We are thus assuming that the cutoff frequency is
directly proportional to the fourth power of the critical
temperature of a given superconductor, and is different for
different superconductive materials. A direct consequence of this
assumption on Beck and Mackey initial model is that the critical
temperature defining the gravitational activity of virtual photons
could also be equal to the usual critical temperature
characterizing the superconductive state. Thus substituting the
critical transition temperature of Niobium, $T_c=9.25K$, into
Equ.(\ref{14}) we find a coupling between the gravitomagnetic
field and the angular velocity of a rotating superconductive
Niobium ring:
\begin{equation}
\chi=1.87\times10^{-4}\label{15}
\end{equation}
which is showing to be extremely close to the above conjectured
coupling based on Cabrera and Tate's measurements of the Cooper
pairs mass in Niobium.
\begin{equation}
\chi=2\frac{\Delta m}{m}=1.84\times10^{-4}\label{16}
\end{equation}

Let us now examine this Coupling for different superconductors
starting with Aluminium and ending with High-$T_c$ superconductors
like YBCO.

\begin{center}
\begin{tabular}{|c|c|c|}
\hline Superconductive material & $T_c [K]$ & $\chi$ \\
\hline $Al$ & $1.18$ & $4.96\times10^{-8}$ \\ \hline $In$ & $3.41$ & $3.46\times10^{-6}$ \\
\hline $Sn$ & $3.72$ & $4.90\times10^{-6}$ \\
\hline $Pb$ & $7.2$ & $6.88\times10^{-5}$ \\
\hline $Nb$ & $9.25$ & $1.87\times10^{-4}$ \\
\hline $High-T_c$ & $79.06$ & $1$ \\
\hline $BSCCO$ & $87.5$ & $1.5$ \\
\hline $YBCO$ & $94$ & $2$ \\ \hline
\end{tabular}
\end{center}
Table 1: Coupling between the gravitomagnetic field and the
angular velocity, $\chi$, of different superconductive materials.
\bigskip

We note that for $YBCO$, with $T_c=94 K$, the gravitomagnetic
London moment transforms exactly into the classical gravitational
Larmor theorem \cite{Mashhoon}:
\begin{equation}
B_g=2\omega\label{16}
\end{equation}
At this point one remark is in order: Our theoretical derivation
presented in this paper strictly speaking holds only for
conventional low-$T_c$ superconductors, because we are using
simple Ginzburg-Landau models and BCS type arguments for both the
superconductor and the dark energy model \cite{Beck4}.

Before concluding it is interesting to note that the
Planck-Einstein scale (involving the fundamental constants:
$\Lambda$, $\hbar$, $c$, $k$, $G$) corresponds to the geometric
mean between cosmological physics (involving the fundamental
constants: $\Lambda$, $\hbar$, $c$, $k$) and high energy particle
physics (involving the fundamental constants: $\hbar$, $c$, $k$,
$G$), just in the scale domain relevant for condensed matter
physics at low temperatures. It is unexpected to have a possible
quantum regime of gravity at this scale
\cite{Padmanabhan2006}\cite{CBCDM}\cite{CDMCB}.

\begin{center}
\begin{tabular}{|c|c|c|c|}
\hline & Einstein scale & Planck-Einstein Scale & Planck scale \\
\hline & $\Lambda$, $\hbar$, c, k & $\Lambda$, $\hbar$, c, k G& c, $\hbar$, k, G \\
\hline Temperature [K] &$T_E=\frac{1}{k}\sqrt{c^2\hbar^2\Lambda}$
& $T_{PE}=\sqrt{T_E T_P}$ & $T_P=\frac{1}{k}\sqrt{\frac{\hbar
c^5}{G}}$ \\
\hline &$2.95\times10^{-55}$ & $60.71$ &
$1.42\times10^{32}$ \\
\hline Time [s] &$t_E=\sqrt{\frac{1}{c^2\Lambda}}$ &
$t_{PE}=\sqrt{t_E t_P}$ & $t_P=\sqrt{\frac{\hbar
G}{c^5}}$ \\
\hline & $2.58\times10^{43}$& $1.26\times10^{-13}$ & $5.38\times10^{-44}$ \\
\hline Length [m] &$l_E=\sqrt{\frac{1}{\Lambda}}$ &
$l_{PE}=\sqrt{l_E l_P}$ & $l_P=\sqrt{\frac{\hbar
G}{c^3}}$ \\
\hline &$8.8\times10^{25}$& $3.77\times10^{-5}$ & $1.61\times10^{-35}$ \\
\hline Mass [Kg] &$M_E=\sqrt{\frac{\hbar^2 \Lambda}{c^2}}$ &
$M_{PE}=\sqrt{M_E M_P}$ & $T_P=\sqrt{\frac{\hbar
c}{G}}$ \\
\hline &$5.53\times10^{-95}$& $9.32\times10^{-39}$ & $2.17\times10^{-8}$ \\
\hline Energy [J] &$E_E=\sqrt{c^2\hbar^2\Lambda}$ &
$E_{PE}=\sqrt{E_E E_P}$ & $E_P=\sqrt{\frac{\hbar
c^5}{G}}$ \\
\hline &$4.07\times10^{-78}$& $8.38\times10^{-22}$ & $1.96\times10^{9}$ \\
\hline Energy density [$J/m^3$]
&$\rho_E=\sqrt{c^2\hbar^2\Lambda^4}$ & $\rho_{PE}=\sqrt{\rho_E
\rho_P}$ & $\rho_P=\sqrt{\frac{c^{14}}{G^4 \hbar^2}}$ \\
\hline &$5.26\times10^{-130}$& $3.73\times10^{-9}$ & $4.6\times10^{113}$ \\
\hline
\end{tabular}
\end{center}
Table 2: Cosmological, versus Planck-Einstein, versus Planck
scales.

Explicitly one has the following formulas at the Planck-Einstein
scale:
\begin{equation}
E_{PE}=kT_{PE}=\Bigg(\frac{c^7\hbar^3 \Lambda}{G}\Bigg)^{1/4}=5.25
[meV]\label{e17}
\end{equation}
\begin{equation}
m_{PE}=\frac{E_{PE}}{c^2}=\Bigg(\frac{\hbar^3
\Lambda}{cG}\Bigg)^{1/4}=9.32\times10^{-39}[Kg]\label{e18}
\end{equation}
\begin{equation}
l_{PE}=\frac{\hbar}{M_{PE}c}=\Bigg(\frac{\hbar G}{c^3
\Lambda}\Bigg)^{1/4}=0.037[mm]\label{e19}
\end{equation}
\begin{equation}
t_{PE}=\frac {l_{PE}}{c}=\Bigg(\frac{\hbar
G}{c^7\Lambda}\Bigg)^{1/4}=1.26\times10^{-13}[s]\label{e20}
\end{equation}
\begin{equation}
\rho_{PE}=\frac{E_{PE}}{l_{PE}^3}=\frac{c^4 \Lambda}{G}=104
[eV/mm^3]
\end{equation}

One readily notices that the numerical values of Planck-Einstein
quantities correspond to typical time, length or energy scales in
superconductor physics, as well as to typical energy scales for
dark energy.

\section{Conclusions}
In conclusion, Table 1 shows that the effective laws of inertia in
superconducting cavities deviate from the laws of classical
mechanics, recovering however the classical regime in the limit of
YBCO cavities. Above, $T_c=94 K$, it is not clear if the classical
gravitational Larmor theorem is affected as indicated in our
model. To investigate this interpretation, from an experimental
point of view, it is recommended to probe classical Coriolis
forces on test masses moving inside rotating superconductive
cavities.

The non-classical laws of inertia in superconductive cavities
arise at the Einstein-Planck regime of gravitation, which appears
to correspond to scales relevant in the domain of condensed matter
physics at low temperatures. It is worth noting that the non
classical rotational inertia exhibited by supersolids could be
another experimental evidence that quantum materials contain a
dark energy density inferior to its cosmological value
\cite{clovis}.

Important questions, related to the proposed approach to dark
energy in superconductors, are open and deserves further
investigation: What couples (decouples) electromagnetic vacuum
energy to (from) gravity? Is Planck-Einstein physics only of
interest to superconductor's physics, or is it also relevant in
other domains of condensed matter physics?

\section{Acknowledgements}
The author would like to thank Prof. Christian Beck, for useful
correspondence on the subject of the present paper, and for
developing, together with Prof. M. C. Mackey, the concept of
electromagnetic dark energy in superconductors.

\end{document}